\documentclass[doublecol]{epl2}

\usepackage{amssymb}
\usepackage[utf8]{inputenc}
\usepackage{slashed}
\usepackage{amsmath}
\usepackage{amsthm}
\usepackage[utf8]{inputenc}
\usepackage{graphicx}
\usepackage{epsfig}
\usepackage{amssymb}
\usepackage{dcolumn}
\usepackage{bm}
\usepackage{color}
\usepackage{hyperref}

\newcommand{\ba}{\begin{array}}
\newcommand{\ea}{\end{array}}
\def\br{\begin{eqnarray}}
\def\er{\end{eqnarray}}
\def\be{\begin{equation}}
\def\ee{\end{equation}}

\def\({\left(}
\def\){\right)}

\def\<{\left\langle}
\def\>{\right\rangle}

\title{Limit on Higgs boson trilinear self-coupling in coupled technicolor models}

\author{A. Doff\inst{1} \and A. A. Natale\inst{2}}

\institute{                    
  \inst{1} Universidade Tecnol\'ogica Federal do Paran\'a - UTFPR - DAFIS
Av Monteiro Lobato Km 04, 84016-210, Ponta Grossa, PR, Brazil\\
  \inst{2} Instituto de F{\'i}sica Te\'orica - UNESP, Rua Dr. Bento T. Ferraz, 271,\\ Bloco II, 01140-070, S\~ao Paulo, SP, Brazil
}

\abstract{
The trilinear self-coupling of the Higgs boson, in a theory in which this boson is composite, is compared to the experimental bound of this quantity obtained by the CMS experiment. In the case of a model where technicolor (TC) is coupled to QCD, we find that the experimental result already constrains the dynamics of the theory, which is represented by an expression of the technifermion self-energy ($\Sigma_{tc}$) typical of technicolor coupled models, and function of the dynamically generated technifermion mass and two other parameters that describe the technifermion dynamical mass momentum dependence. The limits imposed on this dynamics allow us to make a simple determination of pseudo-Goldstone boson masses that appear in these theories, indicating that these bosons may be expected to be quite massive. }

\begin{document}

\maketitle

\section{Introduction}

The Higgs boson discovery by the LHC was one of the major breakthrough of particle physics in the last decades~\cite{atlas,cms}.
With regard to this boson, there is great experimental interest in the possible measurement of its trilinear self-coupling ($\lambda_{HHH}^{SM}$)~\cite{tri}, as well as knowing whether it is a fundamental or composite particle ~\cite{comp}. Any difference between the expected value of the trilinear self-coupling predicted by the standard model (SM) and that of a future measurement of this quantity may indicate a sign of composition or new physics, although the composition or new physics may also arise with the discovery of new particles. In particular, if this boson is composed by new strongly interacting particles, the most discussed signal of a possible dynamical breaking mechanism of the SM gauge symmetry would be the presence of pseudo-Goldstone bosons~\cite{pse}. 

Any limit on the Higgs boson trilinear self-coupling, if this is a composite boson, also means a restriction on the dynamics of the interaction that forms such a boson.
This occurs because the trilinear coupling is directly proportional to the wave function of the composite state and the number of fermions that form that state.
In this work we will compare the trilinear Higgs boson self-coupling computed in the case of technicolor coupled models, showing how the dynamics of the theory is
constrained by the experimental data on this quantity. 

We review how the dynamics of coupled strongly interacting theories are modified compared to an isolated strong interaction theory. In the sequence, based
on the dynamics of these coupled theories, that we assume as QCD and a non-Abelian TC theory coupled by a non-Abelian ETC or GUT, we estimate the
order of the trilinear Higgs boson coupling. With the limits on the dynamics (i.e. technifermion self-energy) originated from the comparison with the
experimental data, we are able to compute pseudo-Goldstone bosons masses in a very simple approximation. The results indicate that these bosons can be
quite massive.

The Lagrangian describing the SM trilinear Higgs boson self-interaction is parameterized as~\cite{tri}
\be
{\cal{L}}^{SM}_{HHH}=\frac{m_H^2}{2v} H^3 \, ,
\label{i1}
\ee
where the SM trilinear coupling with mass dimension is
\be
\lambda^{SM}_{HHH}=\frac{m_H^2}{2v} \, .
\label{i2}
\ee
whose SM expected value 
\be
\lambda_{HHH}^{SM}\equiv \frac{m_H^2}{(2v^2)}=0.129 \,\, .
\label{eq002}
\ee

The Lagrangian describing the observed trilinear Higgs boson self-coupling can be written as
\be
{\cal{L}}_{HHH}= \kappa_\lambda \lambda^{SM}_{HHH} v H^3 \, ,
\label{i3}
\ee
where 
\be
\kappa_\lambda = \frac{\lambda_{HHH}}{\lambda_{HHH}^{SM}} \,\, ,
\label{eq012}
\ee
where $\kappa_\lambda$ is the observed coupling modifier of the trilinear Higgs boson self-coupling.
Recently the CMS Collaboration reported one constraint on the observed coupling $\kappa_\lambda$ at $95\%$ CL~\cite{cms3}
\be
-3.3 < \kappa_\lambda < 8.5 \,\, ,
\label{eq001}
\ee 
This result can already constrain the dynamics of a composite Higgs boson in the context of 
coupled technicolor models~\cite{us1,us2,us3,us4}, and can also be used to determine limits on the possible masses of pseudo-Goldstone bosons.

\section{Dynamics of technicolor coupled models}

Technicolor coupled models are technicolor (TC) models where QCD and TC theories are embedded into a larger gauge group, such that
technifermions and ordinary quarks provide masses to each other~\cite{us1,us2}. In Ref.~\cite{us1} it was verified numerically that
two strongly interacting theories when coupled by another interaction, which could be an extended technicolor theory (ETC) or a
grand unified theory (GUT), have their self-energies (or dynamics) modified when compared to the self-energy of an isolated strong interaction theory.

As the ETC/unified theory should also mediate the interaction of technileptons and ordinary leptons with 
quarks and techniquarks, these fermions also acquire smaller masses than their respective strongly interacting 
partners (i.e. quarks and techniquarks)~\cite{us2}, but as we shall see technileptons also turn out to be quite massive. 

An isolated strong non-Abelian interaction is known to generate a dynamical fermion mass indicated by $\mu$, which is of the order of $\Lambda$, that
is the characteristic scale of the strong interaction. The dynamical fermion self-energy of this strong interaction theory has the following infrared behavior 
(IR)~\cite{mira,rob}
\be
\Sigma (p^2\rightarrow 0)\propto  \mu  \,\, ,
\label{eq1}
\ee
and the ultraviolet behavior (UV) is~\cite{lane2}
\be
\Sigma (p^2\rightarrow \infty )\propto \mu \left( \frac{\mu^2}{p^2} \right) \,\, .
\label{eq2}
\ee

\begin{figure}[t]
\centering
\hspace*{-1cm}\includegraphics[width=1\columnwidth]{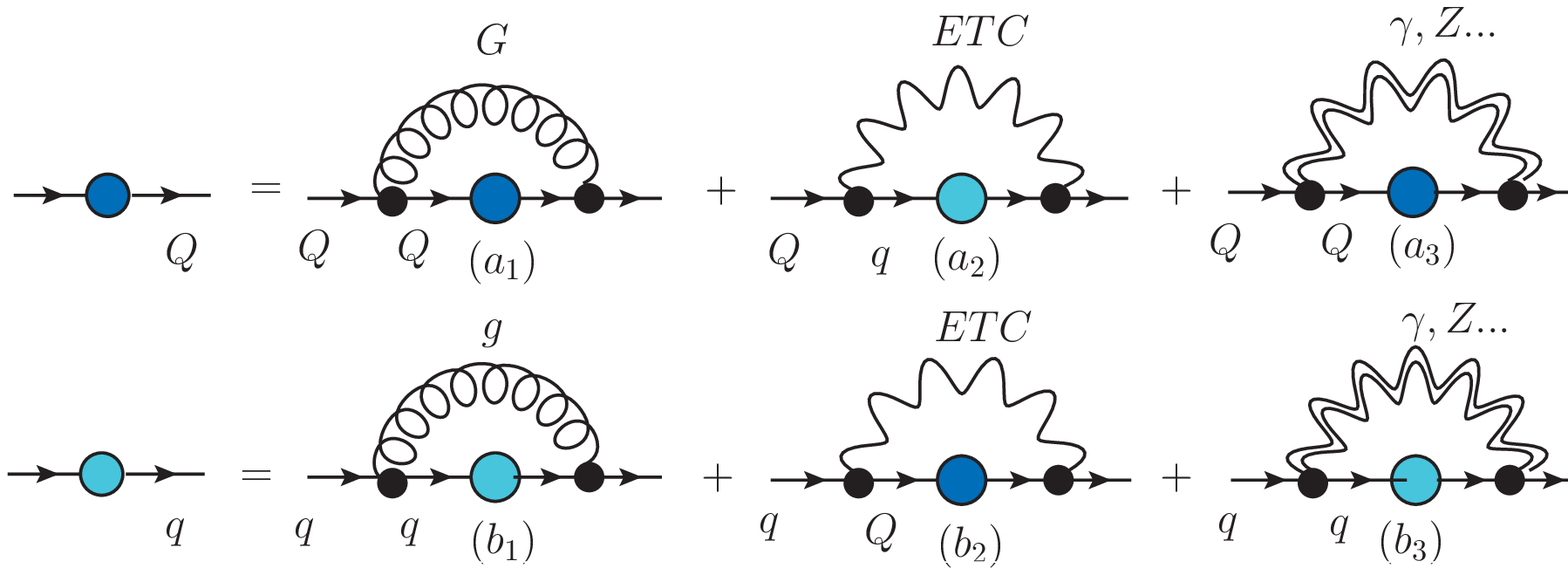}
\vspace*{-8cm}
\caption[dummy0]{The coupled system of SDEs for TC (Q $\equiv$ technifermions) and QCD (q $\equiv$ quarks) including ETC and electroweak or
other corrections. G (g) indicates a technigluon (gluon). }
\label{fig1}
\end{figure}

We can now consider two coupled strong interactions, QCD and TC, through an ETC or GUT theory, where the Schwinger-Dyson equations (SDE) for the
coupled system is depicted in Fig.(1). The IR behavior of both theories is not changed
from the one of Eq.(\ref{eq1}), where now $\mu$ for technifermions will be indicated by $\mu_{tc}$ and for quarks by $\mu_{c}$, respectively the TC and QCD
dynamical fermion masses. However, as shown 
in Ref.\cite{us1,us2}, the effect of QCD and TC to technifermions and quarks is to provide ``bare" masses to each other. We stress this effect, that
is promoted by the second diagram
in the SDE of Fig.(1) for technifermions. Actually, the effect
of this diagram is exactly to change the boundary conditions of the SDE in the differential form, just as it would have if we had introduced a bare mass~\cite{us3}.
In this case the UV behavior of the dynamical self-energy with a ``bare" mass $\mu_0$ is 
given by~\cite{lane2}
\be
\Sigma (p^2\rightarrow \infty )\propto \mu_0 \left[\ln\left( \frac{p^2}{\Lambda^2} \right)\right]^{-\gamma} \,\, .
\label{eq3}
\ee
where $\gamma$ for a $SU(N)$ non-Abelian gauge theory with fermions in the fundamental representation is
\be
\gamma = \frac{3(N^2-1)}{2N(11N-2n_f)} \,\, ,
\label{eq4}
\ee
and $\Lambda$ is the characteristic scale of the theory.
The logarithmic behavior of Eq.(\ref{eq3}) is connected to the running of the non-Abelian gauge coupling constant.

Going back to the coupled SDE system we can notice that the IR behavior of the technifermion self-energy is still proportional to $\mu_{tc}$, as long
as we assume no other new strong interaction above the TC scale, and the technifermion bare masses generated by QCD are very small when compared
to $\mu_{tc}$. 
The actual TC self-energy UV behavior is a combination of a $1/p^2$ component
typical of an isolated TC theory, with the UV logarithmic behavior given by Eq.(\ref{eq3}) as soon as we have momenta larger than $\mu^2_{tc}$, characterized
by the domination of the QCD diagram to the dynamical technifermion mass. Therefore, the full TC dynamical self-energy can be roughly described 
by
\be
\Sigma_{tc} (p^2)\approx \mu_{tc} \left[ 1+ \delta_1 \ln\left[(p^2+\mu_{tc}^2)/\mu_{tc}^2 \right] \right]^{-\delta_2} \,,
\label{eq5}
\ee
Eq.(\ref{eq5}) is the simplest interpolation of the numerical result of Ref.~\cite{us1}, describing the infrared (IR) dynamical mass equal
to $\mu_{tc}$ (also proportional to the technicolor characteristic scale), and a logarithmic decreasing function of the momentum in the ultraviolet (UV) region originated by another (QCD, for instance) strong interaction. 
It is clear that in the IR region the logarithmic term of Eq.(\ref{eq5}) is negligible, and as the momentum increases above $\mu_{tc}$ the logarithmic term controls the UV behavior. 

It is worth to remember that at leading order the fermionic SDE has the same behavior of the scalar Bethe-Salpeter (BS) equation, what
was explicitly shown in Refs.~\cite{ds}. However, the full BS amplitude is subjected to a normalization condition, which, considering Eq.(\ref{eq5}), imposes the following
constraint on $\delta_2$~\cite{lane2,man,chl}
\be
\delta_2 > \frac{1}{2} \, .
\label{eq5a}
\ee
On the other hand, just assuming that $\Sigma (p^2=\mu_{tc}^2)\approx \mu_{tc}$, and that the self-energy starts decreasing smoothly for $p^2>\mu_{tc}^2$, we can assume 
\be
\delta_1\leq 1 \,\, .
\label{eq5b}
\ee
This value is also consistent with the expansion of a dynamical self-energy (e.g. Eq.(\ref{eq3})) at large momentum, where $\delta_1$ would be proportional to the running gauge coupling constant. Ultimately $\delta_1$ may have contributions proportional to $[bg^2]_{si}$ where $b$ and $g$ are respectively the first coefficient of the $\beta$ function and the coupling constant of the strong interaction ($si$) that provides the ``bare" mass to the technifermions 
(see the appendix of Ref.~\cite{cs} to verify the determination of this quantity in the case of an isolated theory).

A consequence of a self-energy like the one of Eq.(\ref{eq5}) is that TC coupled models must incorporate a family symmetry, in such a way that technifermions couple at leading order only to the third ordinary fermion family, whereas the first fermionic family will be coupled at leading order only to QCD~\cite{us1,us3,us4}, i.e. 
the mass hierarchy between different ordinary
fermionic generations can only be obtained through the introduction of a family (or horizontal) symmetry, as described in Refs.\cite{us1,us3,us4}. We will
not touch these aspects here, and in the following we just verify consequences of Eq.(\ref{eq5}) for the trilinear Higgs boson self-coupling and
pseudo-Goldstone masses. The result will be compared with the recent experimental constraint on the trilinear Higgs boson coupling~\cite{cms3}.

\section{Trilinear coupling of a composite Higgs boson}

\begin{figure}[t]
\centering
\includegraphics[width=0.5\columnwidth]{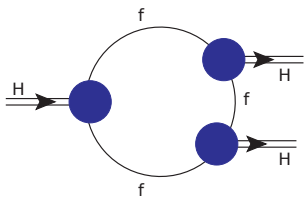}
\caption[dummy0]{ The trilinear composite scalar coupling: The dark (blue) blobs in this figure represent the coupling of composite Higgs bosons(H) to
fermions(f). The double lines represent the composite Higgs bosons. The full diagram is the main contribution to the trilinear Higgs boson self-coupling }
\label{fig2}
\end{figure}
\par 

The trilinear composite scalar coupling is shown in Fig.(2), where the double lines represent the composite Higgs boson, that is coupled to fermions (single line)
through the dark (blue) blobs. 
In the SM the composite scalar boson coupling to fermions (the dark blob) can be determined using Ward identities to be~\cite{so}
\be
G^a(p+q,p)=-\imath \frac{g_W}{2M_W}\left[ \tau^a\Sigma(p)P_R - \Sigma(p+q)\tau^aP_L\right] \, ,
\label{eq6}
\ee
where $P_{R,L}=\frac{1}{2}(1\pm \gamma_5)$, $\tau^a$ is a $SU(2)$ generator, and $\Sigma$ is a matrix of fermionic self-energies in weak-isodoublet space.
At large momenta Eq.(\ref{eq6}) is quite well approximated by $G(p,p)$, and in all situations in which we are interested $\Sigma(p+q)\approx \Sigma(p)$.
Therefore, the coupling given by Eq.(\ref{eq6}) that is dominated by the large momentum running in the loop of Fig.(2) is reduced to
\be
\lambda_{Hff}\equiv G(p,p) \sim - \frac{g_W}{2M_W}\Sigma(p^2) \, .
\label{eq7}
\ee

The loop calculation of Fig.(2), considering Eq.(\ref{eq7}) and $n_F$ technifermions running in that loop, is given by~\cite{us5}
\be
\lambda_{HHH} = \frac{3g^3_W}{64\pi^2} \left(\frac{3n_F}{M^3_W} \right)
\int_0^\infty \frac{\Sigma_{tc}^4(p^2) p^4 dp^2}{(p^2+\Sigma_{tc}^2(p^2))^3} \, .
\label{eq8}
\ee
Note that, apart a dependence on $n_F$, the trilinear coupling is a function of the variables $\delta_1$ and $\delta_2$ shown in Eq.(\ref{eq5}). Of course,
we do also have a dependence on the scale $\mu_{tc}$, but we cannot forget another constraint on the technicolor dynamics that comes from
\be
M_W= \frac{1}{2} g_W F_\pi \,\, ,
\label{eq9}
\ee
where $F_\pi$ is the technipion decay constant, $g_W$ is the electroweak coupling constant, and $F_\pi$ can be calculated through~\cite{ps}
\be
F_\pi^2 = \frac{N}{(2\pi)^2} \int_0^\infty dp^2 \frac{p^2\left[ \Sigma_{tc}^2(p^2)-
\frac{1}{2}p^2\frac{\Sigma_{tc}(p^2)}{dp^2}\Sigma_{tc}(p^2)\right]}{[p^2+\Sigma_{tc}^2(p^2)]^2} \,\, .
\label{eq10}
\ee
Therefore, once the number of technicolors ($N$) and technifermions ($n_F=2n_d$) are specified (where $n_d$ is the number of weakdoublets), the dynamics of the technicolor theory (i.e. $\delta_1$ and $\delta_2$) can be constrained using Eqs.(\ref{eq001}), (\ref{eq5a}), (\ref{eq5b}), (\ref{eq8}), (\ref{eq9}) and (\ref{eq10}). 

Eq.(\ref{eq8}) was already calculated in Ref.~\cite{us5} with a different approximation for Eq.(\ref{eq5}). In that case the self-energy was based on
a possible walking behavior~\cite{ky}, where a certain amount of the $1/p^2$ behavior for this quantity was allowed. Moreover the parameter $\delta_1$ was chosen in an arbitrary
way as $bg^2$, what in a coupled TC scenario does not make sense, due to the many corrections that may contribute to the $\delta_i$ parameters.

\section{Limit on the trilinear coupling}

\begin{figure}[h]
\centering
\includegraphics[width=0.4\columnwidth]{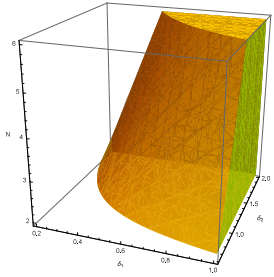}
\caption[dummy0]{$3$-dimensional plot of the technipion decay constant ($F_\pi$) given by Eq.(\ref{eq10}). This quantity is a function of $(\delta_1 , \delta_2 ,  N)$, and
we considered $\mu_{tc}$ in the interval $ 0.5 TeV \leq \mu_{tc} \leq 2TeV$. The yellow region is the allowed one.}
\label{fig3}
\end{figure}

\par In Fig.(3) we present the 3D plot of the technipion decay constant ($F_\pi$) given by Eq.(\ref{eq10}). The plot was generated for $F_{\pi} = v/ \sqrt{3}$, with $v = \sqrt{n_d}F_{\pi} = 246 GeV$ 
assuming  $n_d =3(n_F=6)$ and the following range of technicolor dynamical masses $ 0.5 TeV \leq \mu_{tc} \leq 2TeV$. The dependence of the technipion decay constant on $\mu_{tc}$ is
not appreciable. However, there is a large parameter space for the quantities $(\delta_1 , \delta_2 ,  N)$ that satisfy the experimental $F_\pi$ value. The main relevant fact is the variation
of this quantity with $N$ (the number associated to the technicolor gauge group). For instance, the figure above illustrates that in the region where $N \leq SU(5)_{tc}$, we still have a large volume allowed for  $\delta_1$ and $\delta_2$.  

Considering Eqs.(\ref{eq012}),(\ref{eq5a}), (\ref{eq5b}), (\ref{eq8}) and (\ref{eq9}), in Fig.(4) we present the behavior obtained for Eq.(\ref{eq012}), calculated
assuming the dynamics prescribed in Eq.(\ref{eq5}), $\mu_{tc}= 1TeV$ and $n_f=2$. We also include in the figure the upper limit on the observed coupling 
modifier ($\kappa_\lambda$) 
of the trilinear Higgs boson self-coupling of Ref.~\cite{cms3}, which is indicated
by the dotted-dashed black line.

\begin{figure}[h]
\centering
\includegraphics[width=0.7\columnwidth]{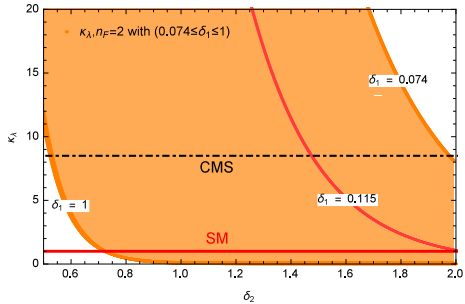}
\caption[dummy0]{The region of allowed $(\delta_1,\delta_2)$ values obtained for the coupling modifier $\kappa_{\lambda}$. 
In this figure we consider  $\mu_{tc}= 1TeV$ and $n_f=2$, furthermore we assume $N=2$ which allows the largest region of parameters bounded by
Eq.(\ref{eq10}). The expected SM value is also indicated by a continuous line.}
\label{fig4}
\end{figure}

\par  In the filled region below the dotted line it is shown the $(\delta_1,\delta_2)$ parameter space allowed by the
experimental constraint on $\kappa_\lambda$, which in this case corresponds to $\delta_1 \geq 0.074$ and $\delta_2 \geq 0.53$. In the Fig.(5) we consider the case where $n_f=4$, which is a little bit more restrictive than the previous one.

\begin{figure}[t]
\centering
\includegraphics[width=0.7\columnwidth]{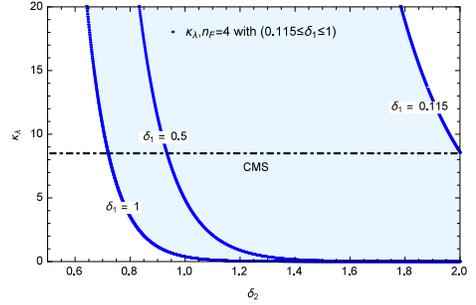}
\caption[dummy0]{ The allowed region of $(\delta_1,\delta_2)$ values obtained for the coupling modifier $\kappa_{\lambda}$. 
We consider again $\mu_{tc}= 1TeV$, $N=2$ and now we set $n_f=4$. }
\label{fig5}
\end{figure}

\par  The case corresponding for $n_f=6$ is described in Fig.(6). Table 1 summarizes the $(\delta_1,\delta_2)$ parameter region allowed  by  the observed 
coupling $\kappa_\lambda$ reported by the CMS experiment, and define the lower limits for $(\delta_1,\delta_2)$. Note that we have not considered $\delta_2$ 
values larger than $2$, which is reasonable if the UV behavior of the TC self-energy is dominated by QCD with $6$ quarks, although other corrections 
to the coupled non-linear SDE system may modify this quantity.

\begin{figure}[h]
\centering
\includegraphics[width=0.7\columnwidth]{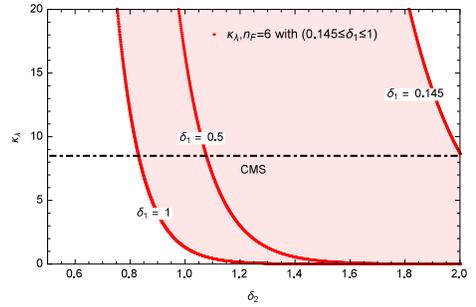}
\caption[dummy0]{The allowed region of $(\delta_1,\delta_2)$ values obtained for the coupling modifier $\kappa_{\lambda}$, 
with  $\mu_{tc}= 1TeV$, $N=2$ and  setting $n_f=6$. }
\label{fig6}
\end{figure}

\begin{table}[t]
\caption{ Parameter region  allowed for the technicolor self-energy dynamics given by Eq.(\ref{eq5}) (i.e. the $\delta_1$ and $\delta_2$ values) constrained by  the observed coupling $\kappa_\lambda$ reported by CMS for $n_f =2,4,6$.}
\label{tab1}
\begin{center}
\begin{tabular}{|c|c|c|c|}
\hline 
$n_F$ &  \hspace*{0.4cm}  $ \delta_1 \geq $  \hspace*{0.4cm} & \hspace*{0.5cm} $ \delta_2 \geq$ \hspace*{0.5cm} & \hspace*{0.2cm} $v/F_{\pi}$ \hspace*{0.2cm} \\ 
\hline
 2  &  $0.074$  & $0.53$ &  $1$  \\  
 4  &  $0.115$  & $0.72$ & $\sqrt{2}$\\ 
 6  &  $0.145$  & $0.83$ &  $\sqrt{3}$  \\ 
\hline 
\end{tabular}
\end{center}
\end{table}

The CMS upper bound on $\kappa_\lambda$
is indicated in the above figures by a dotted-dashed line and is already constraining the dynamics of composite coupled models for the Higgs boson.

We do not expect major changes
in our results in the case of technifermions in higher dimensional representations, because the parameters $\delta_1$ and $\delta_2$ are proportional to
to the product of the Casimir operator of a given representation times the TC coupling constant, and according to the most attractive channel (MAC) hypothesis
the TC chiral symmetry breaking occurs when this product is of $O(1)$ no matter the representation.

\section{Pseudo-Goldstone boson masses}

In technicolor models it is usual to have a large number of pseudo-Goldstone bosons (or technipions) resulting from the chiral symmetry breaking of the
technicolor theory.
In coupled models like the ones discussed in Refs.~\cite{us2} and \cite{us4}, these technipions, besides the ones absorbed by the $W$'s and $Z$ gauge bosons, 
will be of the following type: \\
a) Charged and neutral color singlets, for example,
$$
\bar{U}_iD^i-3\bar{N}E \, ,
$$
$$
\bar{U}_iU^i-\bar{D}_i D^i-3(\bar{N}N-\bar{E}E) \, ,
$$
b) Colored triplets, for example,
$$
\bar{E} U \, ,
$$
c) Colored octets, for example
$$
\bar{U} \frac{\lambda^a}{2} U \, ,
$$
where $\lambda^a$ is a Gell-Mann matrix. The colored triplet and colored octet technipions may be labeled as $\Pi^{(3)}$ and $\Pi^{(8)}$.

\par  Considering the  parameter space of $\delta_1$ and $\delta_2$ values allowed  by  CMS results shown in Table 1, we can discuss what happens with 
the  limits on the masses for the lightest  pseudo-Goldstone bosons expected in the TC coupled scenario when we use the numbers of that table and Eq.(\ref{eq5})
to compute technifermion masses. The heaviest  pseudo-Goldstone bosons carry color once they have large radiative corrections from QCD, while others may have only electroweak corrections to their masses.  In the coupled scenario the lightest technifermion will be the neutral one $(N)$. Apart from TC quantum number the technifermion $N$ has the same quantum numbers of the ordinary neutrino. Its mass appears due to the diagrams of Fig.(7) in models like the ones of 
Ref.~\cite{us2,us4}.

\begin{figure}[ht]
\centering
\hspace*{-1cm}\includegraphics[width=1\columnwidth]{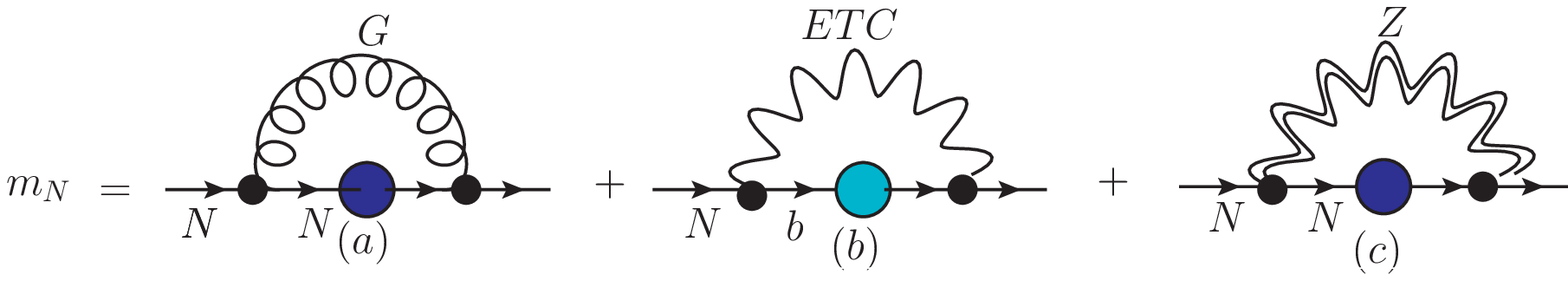}
\vspace*{-10cm}
\caption[dummy0]{ Contributions to the neutral technifermion mass in the coupled TC scheme. }
\label{fig6}
\end{figure}

The diagram (a) of Fig.(7) provides the usual dynamical TC mass to $N$. Remembering that it is the diagram ($a_2$) of Fig.(1) that modifies the running of the $N$ technifermion self-energy,
wich turn out to be logarithmic due to the coupling to the QCD self-energy. The diagram (b) of Fig.(7), in models like the ones of Ref.~\cite{us2,us4}, corresponds to the ETC correction for $m_N$ due to the quark $b$, however, it can be disregarded since $m_b << \mu_ {tc}$. The third diagram of Fig.(7) involves the TC condensate and a weak correction, and this contribution is independent  of any specific ETC model. In a more general scenario,  ETC  gauge bosons can generate corrections similar to that of Fig. (7c), which will not be taken into account in the present work, since we just intend to present simple limits on the spectrum of the lightest pseudo-Goldstone bosons that can eventually be produced in the TC coupled scenario.
\par Considering Eq.(\ref{eq5}), the technilepton ($N$) current mass due to Fig.(7c) can be estimated. The diagram
was calculated at one ETC energy scale $\Lambda_{ETC}\approx \Lambda_{GUT}$ where $\alpha_w \approx \alpha_{ETC} \approx \alpha_{GUT}\approx 0.032$, and the result is
given by

\br 
m_N \approx \frac{3\alpha_w}{4\pi} \frac{\mu_{tc}}{\delta_1\delta_2}  - \frac{3\alpha_w \mu_{tc}}{4\pi}\ln(\frac{M^2_{Z}}{\mu^2_{tc}})
\label{eq11}
\er

\par  Based on this estimate,  assuming the limits described in Table 1,  as well as $\mu_{tc} = 1 TeV$ and $M_{Z} = 91.2 GeV$ we obtain

\br 
&& m_N \approx  231.4 GeV\,\,,\,\,(n_F =2) \nonumber \\ 
&& m_N \approx  151.29 GeV\,\,,\,\,(n_F =4) \nonumber \\ 
&& m_N \approx  100 GeV\,\,,\,\,(n_F =6)
\er 

\noindent  The above results  for $m_N$ follow  from the upper limit on $\kappa_\lambda$ reported by CMS and $\delta_1$ and $\delta_2$ values 
presented in Table 1. These are the $m_N$ masses
obtained in the case of $(2 \leq n_F \leq 6)$. However,  note that for a realistic ETC model, where new interactions including $N$ and ETC bosons are accounted, we 
shall obtain even higher $m_N$ masses. It is important to stress that all other corrections to colored or charged technifermion masses are larger than this one
due to the larger charges and coupling constants (basically changing $\alpha_w$ by $\alpha_s$ and $M_Z$ by a dynamical gluon mass in Eq.(\ref{eq11})).

\par   As neutral technifermions may have masses heavier than $100$ GeV we can determine the mass of the lightest pseudo-Goldstone composed with this neutral particle 
( for instance, $\Pi^N \rightarrow {\bar{N}}\gamma_5\tau^i N$, where $i$ indicate electroweak indexes). This neutral pseudo-Goldstone boson will obtain a mass that may be computed with the help of the  Gell-Mann-Oakes-Renner relation
$$
m_\Pi^{N} \approx \sqrt{m_N \frac{\< {\bar{N}}N\>}{2F_{\Pi}^2} }\, , 
$$
where $\<{ \bar{N}}N\>\approx (\mu_{tc})^3$GeV$^3$ is the TC condensate. However, we
may follow a very simple hypothesis, where the pseudo-Goldstone masses are determined just as the addition of the current masses of their 
constituents~\cite{sca1,sca2}, which was shown to be satisfactory for QCD phenomenology. In this case, supposing that the neutral technipion ($\Pi^N$) is composed just by two $N$ particles we have
\be
m_{\Pi^{N}} \approx 200 \,  - \, 460 \, GeV \, .
\label{f1}
\ee
Notice that we assumed that such neutral boson is solely composed by $N$ technifermions. In general the composition is more complex according to the symmetries
of the TC group, and this neutral boson will also be composed by charged and colored particles increasing the above estimate.

\par Charged and colored technifermions will not only have larger masses than the neutral technifermion, but also more radiative corrections
to their masses, and we can expect even larger masses for colored and charged pseudo-Goldstone bosons.
For instance, following the same hypothesis, the colored triplet and colored octet technipions $\Pi^{(3)}$ and $\Pi^{(8)}$ will obtain masses
\be
m_{\Pi^{(3)}} \approx m_U + m_E \,\, ,
\label{eq61}
\ee
where $m_U$ and $m_E$ are the current masses of the $U$ and $E$ techniquarks. Along the same proposal a simple estimate of the colored octet technipion
of item c) would be 
\be
m_{\Pi^{(8)}} \approx 2 m_U \, .
\label{eq71}
\ee
Changing the weak coupling by the QCD one in the calculation of the $N$ technifermion mass in order to estimate the $U$ and $E$ masses, we can predict $\Pi^{(3)}$ and $\Pi^{(8)}$ masses certainly to be
above $400$ GeV, only with the naive assumption the the strong coupling constant is at least twice the value of the weak one at the TC scale.

\section{Conclusions}

In technicolor coupled models, where TC and QCD are embedded into a large gauge theory, technifermions and ordinary fermions provide bare masses to each other.
In this case the self-energy dynamics of technifermions can be described by Eq.(\ref{eq5}), as verified in Refs.~\cite{us1,us3}.

With the technifermion self-energy given by Eq.(\ref{eq5}) we have computed the trilinear self-coupling of a composite Higgs boson. This calculation is compared
to the recent limits on this coupling obtained by the CMS experiment. The comparison with the experimental data can constrain the trilinear coupling and consequently the dynamics of the TC theory. Once the TC scale ($\mu_{tc}$) is specified we can obtain limits on the variables $\delta_1$ and $\delta_2$ of
Eq.(\ref{eq5}) describing the TC self-energy. Our main result is that the recent experimental data about the trilinear Higgs boson self-coupling is already imposing limits on the TC dynamics, although it is still far from the expected SM value for this quantity. The Higgs boson coupling has been determined
with high precision in the case of heavy fermions, and it would be interesting to verify how the composite wave-function (i.e. self-energy) discussed here is affected by these experimental limits, although in this case the calculation is much more dependent on the ETC/GUT masses and horizontal symmetries necessaries for this type of model.

After obtaining a constraint on the parameters of the TC self-energy for one specific TC scale and number of technifermions we can calculate the technifermion
bare masses. With the values of Table 1, a technicolor mass scale around $1$ TeV, and assuming the simple hypothesis of Refs.~\cite{sca1,sca2}, where the pseudo-Goldstone boson masses are roughly
given by the sum of the particle masses that participate in the boson composition, we can estimate that pseudo-Goldstone boson masses. If these models
are realized in Nature, the pseudo-Goldstone boson masses may be at the order or above $0.5$ TeV.

\acknowledgments

This research  was  partially supported by the Conselho Nacional de Desenvolvimento Cient\'{\i}fico e Tecnol\'ogico (CNPq)
under  grants No 303588/2018-7 (A.A.N.) and 310015/2020-0 (A. D.).

\end{document}